\documentstyle[prd,aps]{revtex}
 
\begin{document} 

\title{On the Stewart-Lyth inverse problem 
\footnote{Published in the Proceedings of III DGFM-SMF Workshop on Gravitation 
and Mathematical-Physics, Le\'on, M\'exico, 2000. }} 
\author{E. Ay\'on-Beato, A. A. Garc\'{\i}a, R. Mansilla$^\dagger$, 
C.A. Terrero-Escalante \\ 
Departamento de F\'{\i}sica, CINVESTAV-IPN\\
Apdo. Postal 14-740, 07000, M\'exico D.F., M\'exico.\\
$^\dagger$ Departamento de F\'{\i}sica Te\'orica, Instituto de F\'{\i}sica, 
UNAM,\\
Apdo. Postal 20-364, 01000, M\'exico D.F., M\'exico.\\
}

\maketitle 
\begin{abstract} 
In this paper the Stewart-Lyth inverse problem was rewritten using the comoving
scale as the basic parameter. It is shown that some information on the 
inflaton potential can be obtained taking into account only the scalar power
spectrum.
\end{abstract} 
\section{Introduction}
During inflation, the inflaton and graviton fields undergo quantum 
fluctuations. These fluctuations generated the seeds for the large-scale 
structure of the
 Universe and left their imprints in the anisotropies 
currently observed in the Cosmic Microwave Background Radiation (CMBR) 
(for detailed 
explanation see Refs.~\cite{Linde} and \cite{KT}).
 
The Stewart and Lyth equations \cite{StLy} are the state of the art for 
predicting the anisotropies for most inflationary models \cite{Lid4}. 
Anticipating the near future launching of satellites able to measure the CMBR 
anisotropies with unprecedented accuracy, we would like to find out whether 
the inflationary potential can be obtained from the Stewart-Lyth equations by 
using as input observational information about the density perturbations and 
gravitational waves spectra.

\section{The Stewart-Lyth equations} 
The theoretical frame for the Stewart-Lyth calculation is the flat 
Friedmann-Robertson-Walker universe containing a single scalar field 
equivalent to a perfect fluid with equations of motions given by
\begin{eqnarray} 
\label{Friedmann} 
H^2 &=& \frac{\kappa}{3}\left(\frac{\dot{\phi}^2}{2} + V(\phi)\right),\\
\label{mass} 
\ddot{\phi} &+& 3H\dot{\phi} = -V^\prime(\phi),
\end{eqnarray}
where $\phi$ is the inflaton, $V(\phi)$ is the inflationary potential, 
$H=\dot{a}/a$ is the Hubble parameter, $a$ is the scale factor, dot and prime 
mean the derivatives with respect to cosmic time and $\phi$ respectively, 
$\kappa = 8\pi/m_{\rm Pl}^2$ is the Einstein constant and $m_{\rm Pl}$ is the 
Planck mass.

In this frame the first three slow-roll parameters are respectively defined 
as \cite{Lid5}(we will write them here as in Ref.~\cite{Lid3a} but using 
$\kappa$):
\begin{eqnarray}
\label{SRP1}  
\epsilon(\phi) &\equiv& 3\frac{\dot{\phi}^2}{2}\left[\frac{\dot{\phi}^2}{2} 
+ V(\phi)\right]^{-1} = \frac{2}{\kappa}\left[\frac{H^\prime}{H}\right]^2,\\
\label{SRP2} 
\eta(\phi) &\equiv& -\frac{\ddot{\phi}}{H\dot{\phi}} = \epsilon 
- \frac{\epsilon^\prime}{\sqrt{2\kappa\epsilon}},\\
\label{SRP3} 
\xi(\phi) &\equiv& \left[\epsilon\eta 
- \left(\frac{2}{\kappa}\epsilon\right)^{\frac{1}{2}}
\eta^\prime\right]^{\frac{1}{2}}.
\end{eqnarray}
Power-law inflation \cite{Lucchin} is a particular case of inflationary model 
where:
\begin{eqnarray}
\label{PL1}
a(t) &\propto& t^p,\\  
\label{PL2} 
H(\phi) &\propto& \exp\left(-\sqrt{\kappa/2p}\,\phi\right),\\
\label{PL3} 
V(\phi) &\propto& \exp\left(-\sqrt{2\kappa/p}\,\phi\right),
\end{eqnarray}
with $p$ a positive constant. It follows from Eqs.~(\ref{SRP1}), (\ref{SRP2}) 
and (\ref{SRP3}) that in this case the slow-roll parameters are constant and 
equal each other.

Exact expressions for the asymptotic scalar (density perturbations) and 
tensorial (gravitational waves) power spectra for the case of power-law 
inflation was correspondingly derived by Lyth and Stewart \cite{LySt} and 
Abbot and Wise \cite{Abbot}.
Assuming that the deviation of the higher slow-roll parameters from $\epsilon$ 
is small (power-law approximation) and that $\epsilon$ is small with respect 
to unity (slow-roll approximation), Stewart and Lyth \cite{StLy} derived 
next-to-leading order expressions for both spectra. In terms of the spectral 
indexes, these expressions are
\begin{eqnarray}
\label{StLythNs}  
1-n_s(k) &\simeq& 4\epsilon - 2\eta + 8(C+1)\epsilon^2 - (10C+6)\epsilon\eta + 2C\xi^2,\\
\label{StLythNt} 
n_T(k) &\simeq& -2\epsilon\left[1 + (2C+3)\epsilon - 2(C+1)\eta \right],
\end{eqnarray}
where the notation is that of Ref.~\cite{Lid3a}, $n_s(k)$ and $n_T(k)$ are the 
scalar and tensorial spectral indices respectively, $k$ is the wave number of 
the comoving scale and $C \approx -0.73$ is a constant related to the Euler 
constant originating in the expansion of the Gamma function. The symbol 
$\simeq$ is used to indicate that these equations were obtained using the 
power-law and slow-roll approximations. Hereafter we will use the equal sign 
in our calculations, but the mean of approximation should be added whenever it 
applies.

Giving an expression for the inflaton potential the definitions (\ref{SRP1}), 
(\ref{SRP2}) and (\ref{SRP3}) should be substituted in the Stewart-Lyth 
equations (\ref{StLythNs}) and (\ref{StLythNt}) to obtain the scale-dependent 
spectral indexes. We should note that the spectral indexes are directly 
related with the power spectra (for details see Ref.~\cite{Lid3a}).

\section{The inverse problem}
If we denote $T \equiv \dot{\phi}^2/2$ and use the definitions (\ref{SRP1}), 
(\ref{SRP2}), (\ref{SRP3}) and the equations (\ref{Friedmann}), (\ref{mass}) 
then we obtain in a straightforward fashion:
\begin{eqnarray}
\label{SRP1T}  
\epsilon &=& 3\frac{T}{T+V} = \frac{\kappa T}{H^2},\\
\label{SRP2T} 
\eta &=&  \kappa \frac{dT}{dH^2},\\
\label{SRP3T} 
\xi^2 &=& \kappa\epsilon\frac{dT}{dH^2} + 2\kappa\epsilon H^2\frac{d^2T}{d(H^2)^2}.
\end{eqnarray}
Denoting $\tau \equiv \ln \, H^2$, $\delta(k) \equiv n_T(k)/2$ and 
$\Delta(k) \equiv [n_s(k) - 1]/2$, substituting expressions (\ref{SRP1T}), 
(\ref{SRP2T}) 
and (\ref{SRP3T}) in equations (\ref{StLythNs}) and (\ref{StLythNt}), and 
rewriting the indexes equations in terms of the first slow-roll parameter 
$\epsilon$ and its derivatives with respect to $\tau$ 
($\hat{\epsilon} \equiv d\epsilon/d\tau$ and 
$\hat{\hat{\epsilon}} \equiv d^2\epsilon/d\tau^2$ ) 
the equations (\ref{StLythNs}) and (\ref{StLythNt}) become:
\begin{eqnarray}
\label{MSch1}
2C\epsilon\hat{\hat{\epsilon}} - (2C+3)\epsilon\hat{\epsilon} - \hat{\epsilon} + \epsilon^2+\epsilon + \Delta &=& 0,\\
\label{MSch2}
2(C+1)\epsilon\hat{\epsilon} - \epsilon^2-\epsilon - \delta &=& 0.
\end{eqnarray}
The {\it{Stewart-Lyth inverse problem}} consists of, giving expressions for 
the spectral indexes, to solve equations (\ref{MSch1}) and (\ref{MSch2}) for 
$\epsilon$ and to find the corresponding inflaton potential.

The second order non-linear differential equation (\ref{MSch1}) was first 
introduced in Ref.~\cite{SchMi} but their approach for the potential 
reconstruction 
is incomplete. First of all, they considered $\Delta(k)$ as a constant. 
Moreover, in Ref.~\cite{SchMi} authors did not take into account the equation 
for the 
tensorial spectral index (\ref{MSch2}).

To proceed with the analysis of Eqs.~(\ref{MSch1}) and (\ref{MSch2}) we note 
that 
in these equations $\epsilon$ and its derivatives depend on $\tau$ while 
$\Delta$ and $\delta$ explicitly depend on $k$. So, we shall rewrite 
Eqs.~(\ref{MSch1}) and (\ref{MSch2}) in terms of $k$. We choose $k$ because 
this is 
the parameter related to observations. Using the equation derived in 
Ref.~\cite{Lid3a} that relates the inflaton value with the corresponding scale,
\begin{equation}
\frac{d\ln k}{d\phi}=\frac{\kappa}{2}\frac{H}{H^\prime}\left(\epsilon-1\right),
\label{k2phi} 
\end{equation}
after conversion we obtain
\begin{eqnarray}
\label{Eqk1}
\frac{C(\epsilon-1)^2}{2\epsilon}\tilde{\tilde{\epsilon}} 
+ \frac{C(\epsilon-1)}{2}\left(\frac{\tilde{\epsilon}}{\epsilon}\right)^2 
- \left[(2C+3)\epsilon + 1\right]\frac{\epsilon-1}{2\epsilon}\tilde{\epsilon}
+\epsilon^2+\epsilon+\Delta&=&0,\\
\label{Eqk2}
(C+1)(\epsilon-1)\tilde{\epsilon}-\epsilon^2-\epsilon-\delta&=&0.
\end{eqnarray}
where $\tilde{\epsilon} \equiv d\epsilon/d\ln k$ and $\tilde{\tilde{\epsilon}} 
\equiv d^2\epsilon/d(\ln k)^2$.
Differentiating Eq.~(\ref{Eqk2}) with respect to $\ln k$ we can replace the 
expressions for $\tilde{\epsilon}$ and $\tilde{\tilde{\epsilon}}$ obtained 
from this equation into Eq.~(\ref{Eqk1}). After that, we obtain the following 
algebraic equation for $\epsilon$:
\begin{equation}
\label{Eq4degree}
\epsilon(k)^4+P\epsilon(k)^3+Q(\tilde{\delta},\delta, \Delta)\epsilon(k)^2+R(\tilde{\delta},\delta)\epsilon(k)+S(\delta)=0,
\end{equation}
where
\begin{eqnarray}
\nonumber
P&=&C+2,\\
\nonumber
Q(\tilde{\delta},\delta, \Delta)&=&-(C+1)\left [C\tilde{\delta}-(2C+3)\delta+2(C+1)\Delta-1 \right ],\\
\nonumber
R(\tilde{\delta},\delta)&=&(C+1)C\tilde{\delta}+(2C+1)\delta,\\
S(\delta)&=&C\delta^2.
\nonumber
\end{eqnarray}
Using observational data of the spectra as input in Eq.~(\ref{Eq4degree}) one 
can 
solve the Stewart-Lyth inverse problem. As an example, one can see that for 
$\Delta$ and $\delta$ constants the unique solution is just power-law 
inflation, i.e., $\epsilon=\rm constant$.

The roots of Eq.~(\ref{Eq4degree}) can be readily calculated but they are very 
complicated expressions not necessary for further analysis. Instead, we can 
see that the roots will have the form:
\begin{equation}
\label{Roots4}
\epsilon_i=\epsilon_i(\Delta, \delta, \tilde{\delta}),
\end{equation}
where $i=1,\dots,4$. From set (\ref{Roots4}), those $j$ roots with 
$0<\epsilon \leq 1$ (in, at least, a finite range of the scale) should be 
chosen. Then, $j \leq i$.

Differentiating Eq.~(\ref{Roots4}) with respect to $\ln k$ and substituting in 
Eq.~(\ref{Eqk2}) we will obtain equations of the form:
\begin{equation}
\label{F4}
{\cal{F}}_j(\Delta, \tilde{\Delta}, \delta, \tilde{\delta},
\tilde{\tilde{\delta}} )=0.
\end{equation}
We can see that although Eqs.~(\ref{F4}) can not be solved for both spectra, 
to obtain explicit expressions for $\epsilon_j$ it is, in principle, 
sufficient to have information about $\Delta(k)$ or $\delta(k)$, or a 
relationship between them.
If a functional form for one of the spectra is proposed and the other one 
determined from Eq.~(\ref{F4}), the corresponding inflaton potential can be 
determined uniquely.
Having explicit expressions of $\epsilon_j$ as functions of $k$ equations 
(\ref{SRP1}) and (\ref{k2phi}) can be used to obtain $H_j(k)$:
\begin{equation}
H_j(k)=H_{0j}\exp\left[\int{\frac{\epsilon_j(k)}{\epsilon_j(k)-1}d\ln k}\right],
\label{Hk}
\end{equation}
where $H_{0j}$ are integration constants.
Then, using the equations of motion (\ref{Friedmann}) and (\ref{mass}), and 
definition (\ref{SRP1T}) the potential as a function of scale is obtained:
\begin{equation}
V_j(k)=\frac{H_j^2(k)}{\kappa}(3-\epsilon_j(k)).
\label{Vk}
\end{equation}
On the other hand, using definitions (\ref{SRP1}) and (\ref{k2phi}) the scalar 
field as a function of the scale is given by ($\phi_{0j}$ are integration 
constants):
\begin{equation}
\phi_j(k)=\phi_{0j}-\sqrt{\frac{2}{\kappa}}\int{\frac{\sqrt{\epsilon_j(k)}}{\epsilon_j(k)-1}d\ln k}.
\label{Phik}
\end{equation}
Finally, the inflationary potential as function of the inflaton could be given 
parametrically
\begin{equation}
V_j(\phi_j)=
\left\{\begin{array}{ll} 
\phi_j(k)\\
V_j(k)  
\end{array}\right.
\label{FVphi}
\end{equation}
The above expression defines the set of functional forms for the inflationary 
potential that are consistent with the slow-roll and power-law approximations 
and the information used as input to solve Eq.~(\ref{F4}).
\section{Summary and Outlook} 
In general, information about the functional form of only one of the spectra 
(or a relationship between them) is sufficient in order to find the inflaton 
potential. In principle, one can take the scalar spectrum derived from 
anisotropies measurements and try to solve the second order differential 
equation that remains for the tensorial spectral index. Nevertheless, 
being the tensorial modes currently difficult to measure, the 
simplest way of solving the Stewart-Lyth inverse problem seems to be to fix 
the spectrum of tensorial perturbations and solve an ordinary first order 
differential equation for the scalar index. Then, one can compare the analytic 
result with the observations.

If an expression for the first slow-roll parameter as a function of the scale 
could be obtained, the corresponding functional form for the inflaton potential
 could be found, at least parametrically. This expression for the potential 
will be constrained by the slow-roll and the power-law approximations and by 
the input information used to solve the above mentioned differential equation.

In forthcoming papers we will explore the feasibility of this approach by 
fixing the functional form of the gravitational waves spectrum with regards to 
the underlying physics and measurements.

\section{Acknowledgments} 
This research is supported in part by the SNI and the CONACyT (M\'exico) 
under project 32138-E.

\end{document}